\documentclass{pic2012}
\def\runauthor{Nicol\`o Cartiglia}
\def\shorttitle{proton-proton total cross section}
\usepackage{wrapfig}
\begin{document}

\title{Measurement of the proton-proton total cross section  at 2, 7, 8 and 57 TeV}

\author{Nicol\`o Cartiglia}

\address{INFN \\
Via Pietro Giuria 1, 10125 Torino, Italia }

\maketitle

\abstracts{The measurement of the total $pp$ cross section and its various sub-components (elastic, inelastic and diffractive) is a very powerful tool to understand the proton macro structure and fundamental QCD dynamics. In this  contribution I first provide a theoretical introduction to the topic, then a summary of the experimental techniques and finally I review the new results from AUGER and LHC experiments.}

\section{Theoretical framework} 
\begin{figure}[h]
\begin{center}
 \resizebox{10cm}{!}{\includegraphics{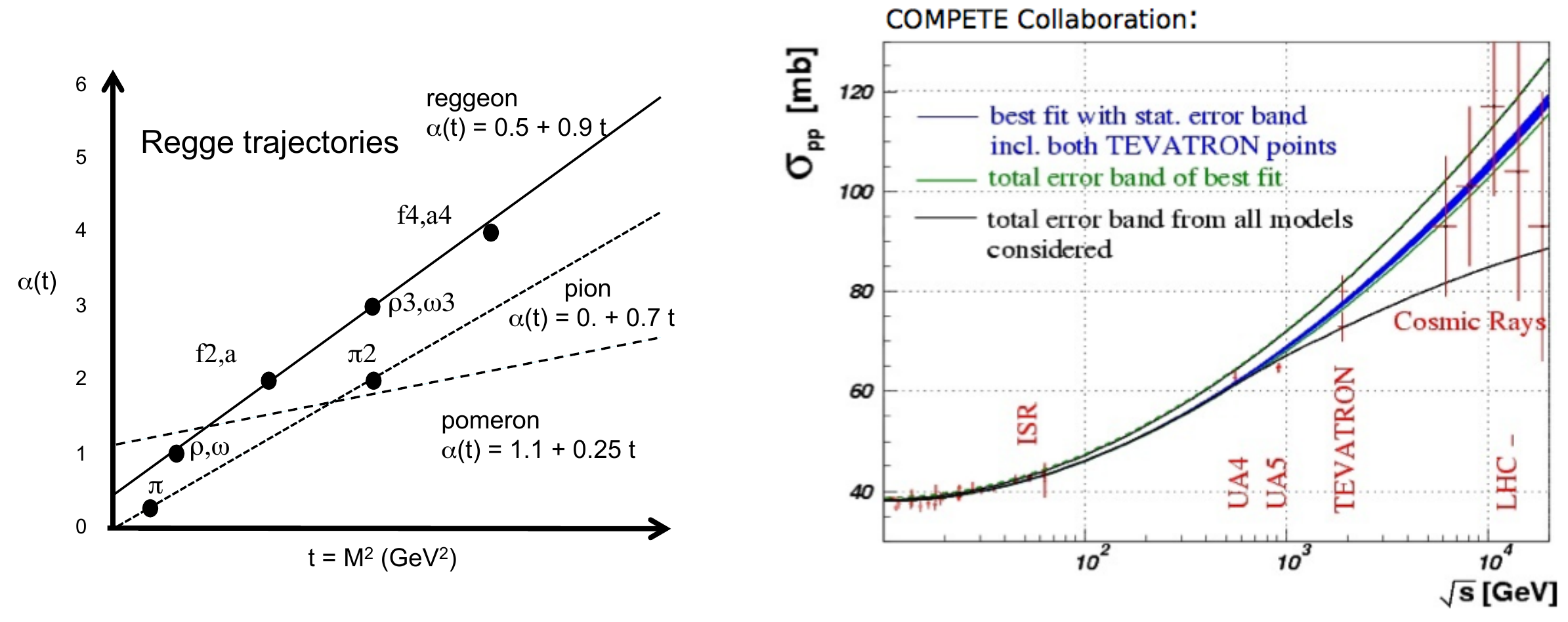}}
\end{center}
\caption{ Left pane: Example of Regge trajectories in the complex angular moment (J) - t plane. Right pane: Evolution of the value of total $pp$ cross section as a function of the center-of-mass energy  as predicted by the COMPETE collaboration. The darkest band is the fit that has the best $\chi^2/DOF$ using pre-LHC points.}
\label{fig:regge_compete}
\end{figure}

Regge theory is the theoretical framework that is used to study many soft QCD processes such as diffraction and the total cross section. The first main feature of the Regge pole model is  the existence, in the complex angular moment $J$ - $t$ plane, where $t$ is  the 4-momentum transfer squared,  of trajectories $\alpha(t) = \alpha + \alpha\prime t$ such that, whenever $t = m^2$ (where $m$ is the mass of a particle in the trajectory), then $\alpha(t)$ correspond to the spin of the particle, Figure~\ref{fig:regge_compete} (left pane).  
\vspace{-0.5cm}
\noindent These trajectories can be understood as group of particles that are exchanged together, i.e., referring to Figure~\ref{fig:regge_compete} (left pane), in the scattering process  $p \pi^o \rightarrow n \pi^+ $ not just the $\rho$ particle is exchanged, but all particles on the $\rho$ trajectory. The second main feature of the Regge pole model is the relation between exchanged trajectories and high-energy behaviour: the contribution to the total cross section from a given trajectory is given by $\sigma(s)  \propto Im A(s, t = 0) \sim s^{\alpha-1}$,  where $Im A(s, t = 0)$ is the imaginary part of the scattering amplitude computed at $ t = 0 $ GeV and  $\alpha$ is the intercept of the exchanged trajectory. The equation of the trajectory indicates a very important feature: if the intercept is lower than one, the contribution of a trajectory to the total cross section decreases as a function of increasing center-of-mass energy. All the trajectories that are formed by known particles have intercepts lower than one and therefore provide a decreasing  contribution to the cross sections. This prediction is, however, not supported by the experimental points: following an initial decrease, the value of the cross section rises as the center-of-mass energy grows, Figure~\ref{fig:regge_compete} (Right pane). This problem  is solved by introducing a new trajectory with an intercept larger than one: the pomeron trajectory, shown also on Figure~\ref{fig:regge_compete} (left pane). The exchanged particles (poles) on the reggeon and pion trajectories offer guidance on how to write the scattering amplitude $A(s, t)$, however this is not the case for the pomeron trajectory, as it has no particles on it. The possible contributions of the pomeron trajectory to the total cross section have different analytic forms depending on the type of diagram considered: 
\begin{eqnarray}
\label{eq:pom1}
\sigma(s)  \propto Im A(s, t = 0) \sim s^{\alpha-1}, \\
\label{eq:pom2}
\sigma(s)  \propto Im A(s, t = 0) \sim ln(s), \\
\label{eq:pom3}
\sigma(s)  \propto Im A(s, t = 0) \sim ln^2(s), 
\end{eqnarray}
\noindent where Equation~\ref{eq:pom1} is for a simple poles type of exchange, while Equation~\ref{eq:pom2} and Equation~\ref{eq:pom3} are for more complicate processes. The various diagrams have been analyzed by the COMPETE~\cite{Nicolescu:2001um}\cite{Cudell:2001ma} collaboration, which has produced a prediction for the evolution of the  value of the total $pp$ cross section as a function of center-of-mass energy, Figure~\ref{fig:regge_compete} (Right pane). The three most common parametrizations of the cross section are:
\begin{eqnarray}
\label{eq:cross1}
\sigma(s)  =  c_1 + c_2 * s^{-0.5} + c_3 * s^{0.08}, \\
\label{eq:cross2}
\sigma(s)  =  c_1 + c_2 * s^{-0.5} + c_3 * ln^2(s), \\
\label{eq:cross3}
\sigma(s)  =  c_1 + c_2 * ln(s) + c_3 * ln^2(s).
\end{eqnarray}
\noindent These studies find that the analytic form that fits the low energy data points better  is Equation~\ref{eq:cross2}  and their best  {\it pre-LHC} predictions are: $\sigma_{Tot}(7 \; TeV) = 98 \pm 5$ mb;  $\sigma_{Tot}(8\; TeV) = 101 \pm 5$ mb;  $\sigma_{Tot}(14 \; TeV) = 111.5 \pm 5$ mb. 
\noindent Following the experimental discovery at HERA of the steep rise of the gluon distribution as a function of center-of-mass energy, the predictions for the total cross section were also updated. In particular, Equation~\ref{eq:cross1} was modified introducing a second simple pole, the so called {\it hard pomeron}~\cite{Landshoff:2007uk} \cite{Cudell:2009bx}: $\sigma(s)  =  c_1 + c_2 * s^{-0.5} + c_3 * s^{0.067} + c_4 * s^{0.45}$.

\begin{figure}[h]
\begin{center}
 \resizebox{10cm}{!}{\includegraphics{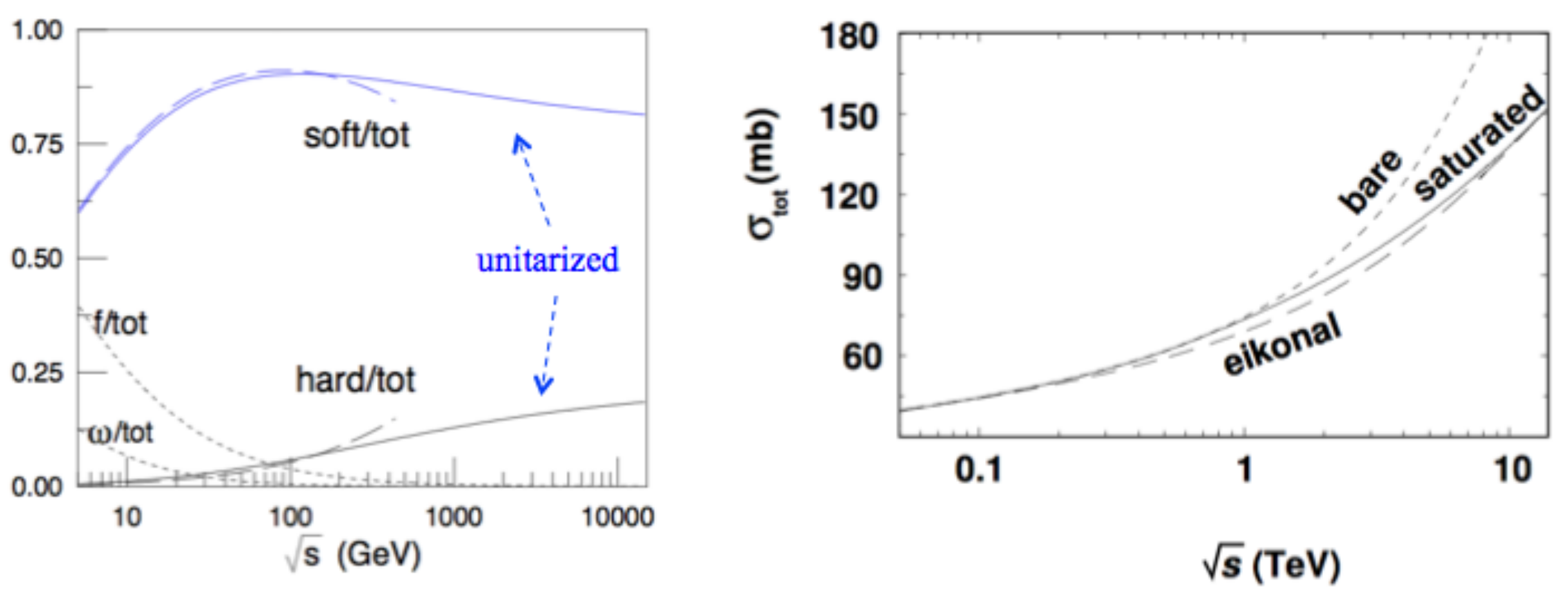}}
\end{center}
\caption{ Left pane: Various contributions to the cross section as a function of the center-of-mass energy. 
Right pane: Predictions of the two-pomeron model without ({\it bare}) and with ({\it eikonal, saturated}) unitarization effects. Figures from [4] }
\label{fig:softhard}
\end{figure}

\noindent The combined fit using the {\it soft} and {\it hard} pomerons lowers the soft pomeron intercept from $0.08$ to $0.067$ while the hard pomeron intercept value of $0.45$ is perfectly compatible with HERA results. Figure~\ref{fig:softhard} (left pane) shows the various contributions to the cross section as a function of the center-of-mass energy. A feature common to many models is the continuous rise of the cross section with energy that eventually leads to the violation of unitarity. In particular, the value of the total cross section has to respect the Froissart-Martin bound: $ \sigma_{Tot}(s) < \frac{\pi}{m_\pi}ln^2(s),$ which at LHC is however  not a real constrain as its value is very large: $\sigma_{Tot}(pp) < 4.3$ barn. The inclusion of this bound and the effects of multiple exchanges in the calculation of  the value of the total cross section is called  {\it unitarization}. The overall effect of unitarization is to reduce the value of the total cross section: Figure~\ref{fig:softhard} (right pane) shows the predictions of the two-pomeron model without ({\it bare}) and with ({\it eikonal, saturated}) unitarization effects. According to~\cite{Cudell:2009bx}, a value $\sigma_{Tot}^{\sqrt{s} = 14 TeV}(pp)$ = 120 - 160 mb would be a clear sign of the two-pomeron model while  a value around 110 mb would be an indication of the $ln^2(s) $ behaviour. It's interesting to note that the prediction from a given parametrization is the outcome of the interplay of its functional expression and the unitarization scheme used.

\begin{figure}[h]
\begin{center}
 \resizebox{10cm}{!}{\includegraphics{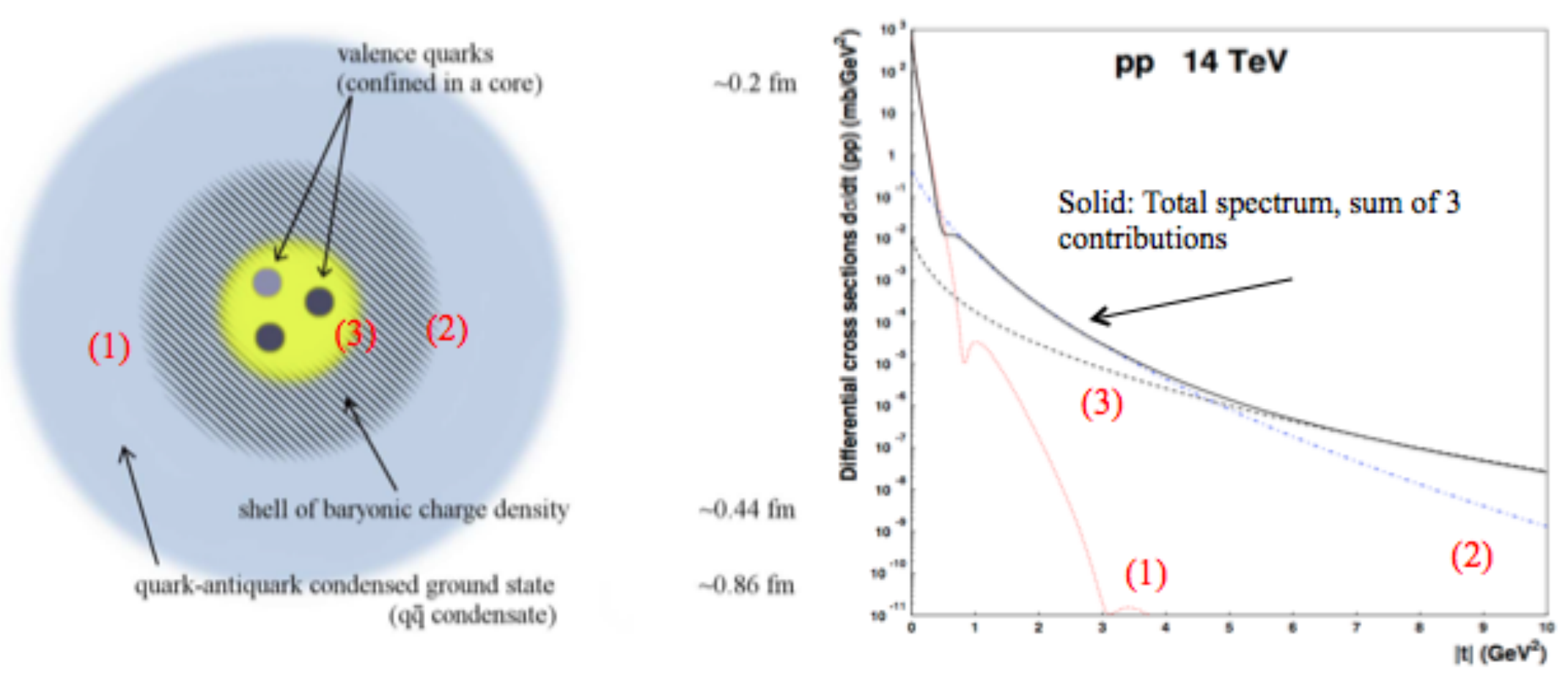}}
\end{center}
\caption{Left pane: sketch of the proton macro structure (see text for details) Right pane: elastic cross section as a function of the 4-momentum transfer $t$. The contributions from the various layers of the proton are shown.}
\label{fig:pstruct}
\end{figure}
\vspace{-0.5cm}
\section{Proton - proton elastic scattering} 
\noindent Elastic scattering, $pp \rightarrow pp$, is a very important process to probe the macro structure of the proton, and it represents roughly one forth of the total cross section.  A sketch of the proton macro structure, following~\cite{Islam}, is shown in Figure~\ref{fig:pstruct} (left pane): the outer corona (1) is composed by $q\bar{q}$ condensates, the middle part is a shell of baryonic charge density (2), while  the valence quarks are confined at the center (3).  

\begin{wrapfigure}{h}{0.4\textwidth}
  \begin{center}
    \includegraphics[width=0.4\textwidth]{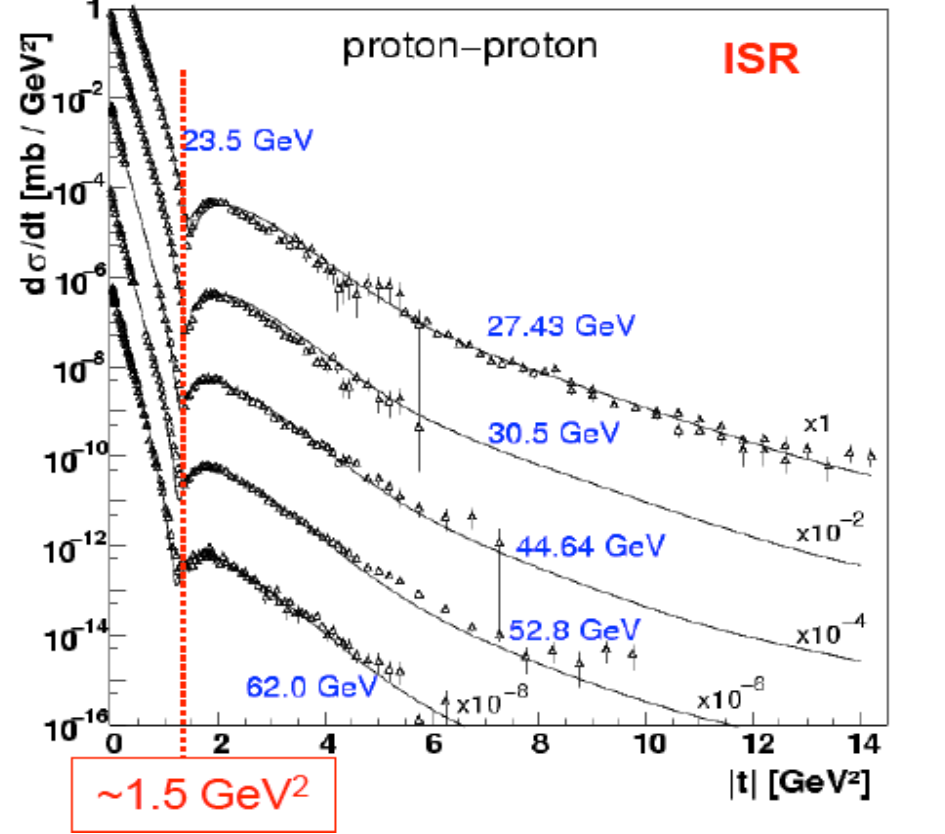}
  \end{center}
    \caption{Differential value of the proton-proton cross section as a function of the 4-momentum transfer squared $t$, for different center-of-mass energies.}
     \label{fig:isr}
\end{wrapfigure}

\noindent Elastic scattering probes the proton at a distance $b$ given by $ b \sim 1/\sqrt{t}$. At low $t$ values, the cross section is well approximated by an exponential  form: $$ \frac{d\sigma}{dt} = A * e^{Bt}, $$ and is largely dominated by the outer corona (1), Figure~\ref{fig:pstruct} (right pane).  At higher values of $t$, the cross section has a more complex form, reflecting the additional contributions from inner layers. At values of $t$ above 4 GeV$^2$ the cross section is dominated by quark-quark elastic scattering (the so called {\it deep elastic scattering}). As the center-of-mass energy grows not only the proton becomes blacker (the cross section increases) but also grows in size: the value of the slope parameter $B$ increases (the so called ``shrinkage of the forward peak''), indicating an average lower values of $t$ and therefore a longer interaction range. The relative importance of the various components described in Figure~\ref{fig:pstruct} (right pane)  changes with energy as shown by the experimental data from ISR, Figure~\ref{fig:isr}, where the  values of the position and depth of the dip decrease with increasing center-of-mass energy. 

\section{Montecarlo models}



The Montecarlo (MC) models commonly used in high-energy and cosmic-rays physics can be approximately divided into two large families. The  MC models in the first group (QGSJET, SIBYLL, PHOJET, EPOS) are based on Regge Field Theory (RFT), and they differ among themselves for the implementation of various aspects of the model parameters. For their focus on soft and forward physics they have been intensively used in cosmic-rays physics and they are a very important tool in the study of the total cross section. These MC models have been extended to also provide predictions for  hard QCD processes. The second group of MC models is based on the calculation of perturbative QCD matrix elements (PYTHIA, HERWIG, SHERPA) and the relative importance and absolute values of  soft processes (total and inelastic cross section, fraction of non-diffractive and  diffractive events) just reflects the chosen internal parametrization.

\section{Topologies of events in $\sigma_{Tot}(pp)$ }


Three main components contribute to the value of $\sigma_{Tot}(pp)$. (i) Elastic scattering $pp \rightarrow pp$: 20-25\% of $\sigma_{Tot} (pp)$. This is a very difficult process to measure and requires dedicated experiments. (ii) Diffractive scattering $pp \rightarrow Xp; \; pp \rightarrow pXp; \; pp \rightarrow XY$: 25 - 35 \% of $\sigma_{Tot}(pp)$. This group includes all processes that are mediated by the exchange of a colour neutral object traditionally called {\it pomeron}. For this reason these events are recognizable by the presence of a large gap in the rapidity distributions of final state particles. This process includes single, double and central diffraction. Low mass diffraction is experimentally  difficult to measure as the hadronic activity produced in the interaction is very small. (iii) Non-diffractive scattering (everything else), this is the easiest part to measure, as normally the events have a large number of particles and can be easily detected. The distinction between elastic and non-elastic scattering is quite obvious, therefore often the results are presented in terms of three components: $\sigma_{Tot}(pp), \; \sigma_{El}(pp)$ and $\sigma_{Inel}(pp)$, while the distinction between diffractive and non-diffractive events is much less straightforward.

\section{Measurement of $\sigma_{Tot}(pp), \;  \sigma_{El}(pp)$, and $\sigma_{Inel }(pp) $ using elastic scattering.}
\vspace{-0.5cm}
\begin{figure}[h]
\begin{center}
 \resizebox{10cm}{!}{\includegraphics{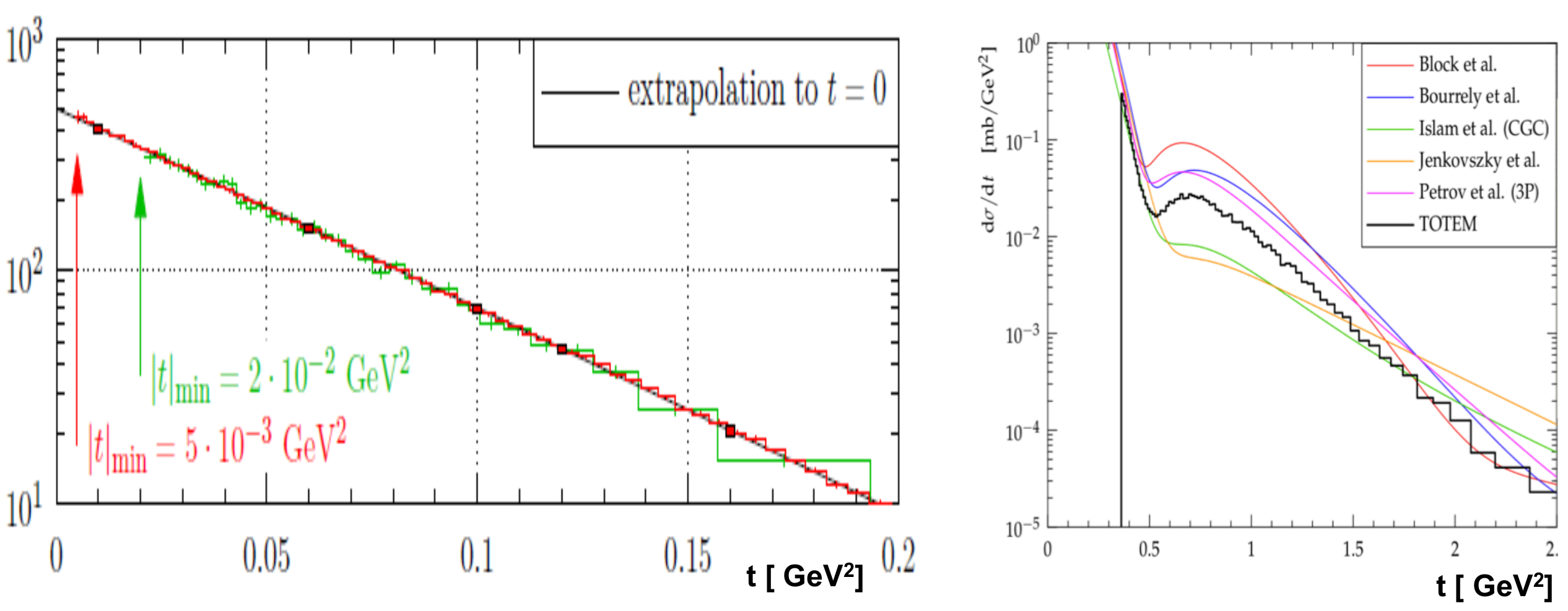}}
\end{center}
\caption{ Left pane: Elastic differential $pp$ cross-section in the very low $t$ range, showing the extrapolation to $t = 0 $ GeV/c. Right pane: Elastic differential $pp$ cross-section in the extended $t$ range as measured by the TOTEM collaboration (black histogram) together with the predictions of several proton models.}
\label{fig:trange}
\end{figure}
\vspace{-0.5cm}
\noindent The TOTEM experiment at the LHC has the capability to detect elastic $pp$ scattering and to measure the differential cross-section for elastic $pp$ scattering as a function of the four-momentum transfer squared $t$ \cite{Antchev:2013gaa}. The results are shown in Figure~\ref{fig:trange} (left pane). In order to derive the elastic and total cross-sections, the extrapolation to the optical point $t = 0$ GeV$^2$ was performed. The elastic data were integrated up to $|t| = 0.415 $ GeV$^2$, where the effect of the larger $|t|$-contributions is small compared to the other uncertainties. The optical theorem can be used to calculate the total $pp$ cross-sections from the value of the $t$ distribution extrapolated to $t = 0$ GeV$^2$:
\begin{equation}
\label{eq:opt}
\sigma^2_{Tot} = \frac{16\pi (\hbar c)^2}{1+\rho^2}\frac{d\sigma_{El}}{dt}|_{t = 0} \;\; \rho = \frac{Re (f_{El})}{Im (f_{El})}|_{t = 0}
\end{equation}
\noindent where the $\rho$ is the ratio of the real to the imaginary part of the forward scattering amplitude. The value of $\rho$ has both been predicted by the COMPETE collaboration ($ \rho^2 \simeq 0.02$) ~\cite{Cudell:2002xe} and measured by TOTEM ($\rho^2 = 0.009 \pm 0.056$)~\cite{Antchev:2013iaa}.  The minimum $t$ value than can be reached using the TOTEM roman pots depends on the LHC optics: as the angular beam spread $\sigma(\theta)$ is inversely correlated with the accelerator parameter $\beta^*$, $\sigma(\theta)  \propto 1/\sqrt{\beta^*}$, it is necessary to have large $\beta^*$ values to achieve a small beam angular spread and therefore be able to  measure low values of $t$. In 2012 the largest value of  $\beta^*$ achieved by the LHC has been $\beta^*= 1000$ m, allowing to detect scattered protons with  $t \simeq 5* 10^{-4}$ GeV$^2$. In the low $t$ range ($ t < 0.4$ GeV$^2$), the TOTEM collaboration measured the value of the slope parameter to be $ B = 19.9 \pm 0.3 $ GeV$^{-2}$, confirming the shrinkage of the forward peak. The TOTEM collaboration has also measured the differential $pp$ distribution in an extended $t$ range: Figure~\ref{fig:trange} (right pane) shows this result together with the prediction of several proton models which do not quite reproduce the data well. Using Equation~\ref{eq:opt}, the value of the $\sigma_{Tot}(pp)$ can be obtained while  the inelastic cross section is obtained by subtraction: $\sigma_{Inel}(pp) = \sigma_{Tot}(pp) - \sigma_{Inel}(pp)$. Using data collected both at $\sqrt{s} =$ 7 and 8 TeV, the TOTEM collaboration has produced a complete set of measurements, shown in Table~\ref{tab:totem}.
\vspace{-0.3cm}
\begin{table}[h]
  \begin{center}
  \begin{tabular}{c c  c c } \hline
Measurement & $\sigma_{Tot}(pp)$ &   $\sigma_{El}(pp)$ &  $\sigma_{Inel}(pp)$ \\ \hline
 $\sqrt{s} = 7 $ TeV &   98.6 $\pm$ 2.8 mb  &  25.4 $\pm$ 1.1 mb & 73.1 $\pm$ 1.3 mb \\
 $\sqrt{s} = 8 $ TeV &  101.7 $\pm$ 2.9 mb &  27.1 $\pm$ 1.4 mb & 74.7 $\pm$ 1.7 mb\\ 
\hline
\end{tabular}
\caption{Values of the total, elastic and inelastic $pp$ cross section at $\sqrt{s} = $ 7 and 8 TeV as measured by the TOTEM collaboration}
\label{tab:totem}
\end{center}
\end{table}
\vspace{-0.5cm}
\section{Measurements of sub-parts of   $\sigma_{Inel}(pp)$ and extrapolation to  $\sigma_{Inel}(pp)$ and $\sigma_{Tot}(pp)$: cosmic-rays and collider experiments}

\noindent As outlined above, the complete measurement of all processes that compose  $\sigma_{Inel}(pp)$ is very difficult and currently no experiment is able to do it. Both cosmic-rays and collider experiments can directly measure only parts of  $\sigma_{Inel}(pp)$ and therefore need to use MC or analytical models if they want to provide an estimate of the full values of $\sigma_{Inel}(pp)$. 

\section{Cosmic-rays experiments}
\begin{figure}[h]
\begin{center}
 \resizebox{11cm}{!}{\includegraphics{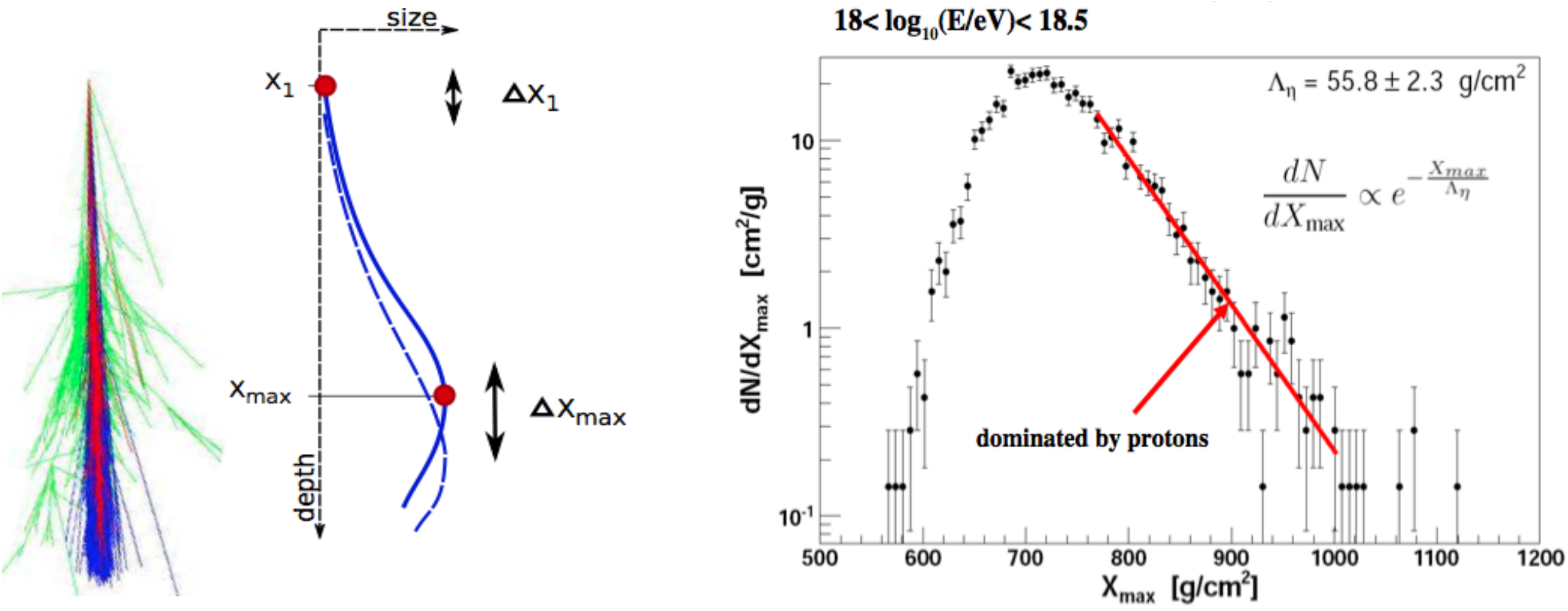}}
\end{center}
\caption{Left pane: Sketch of a cosmic-ray shower development. $x_1$ indicates the position of impact.
Right pane: Position of shower maximum as a function of depth in the atmosphere as measured by the AUGER collaboration. }
 \label{fig:cos12}
\end{figure}

\noindent Cosmic-rays experiments measure the interaction of the primary particle with a nucleus in the atmosphere via the secondary particles, generated in the hadronic shower, that reach ground level. This method is bound to carry significant uncertainties as the measured quantities are only indirectly related to the primary scattering event: the cosmic-rays flux composition, the atmospheric molecular mixture, the modelling of the hadronic shower, and the limited detector acceptance concur to the measurement uncertainties.  The value of $\sigma_{Inel}(p-air)$  directly influence  the distance $x$ that the primary particle travels in air before interacting, Figure~\ref{fig:cos12} (left pane): lower values of the cross section move  the point of interaction $x_1$ deeper into the atmosphere.   There are two main methods to reconstruct the $x_1$ position. (i) $\frac{N_e}{N_\mu}$: the ratio of the number of electrons to the number of muons is related to the shower length. By measuring this number together with the shower direction, the position of $x_1$ can be determined. This method relies on  MC models to simulate the shower development and to correctly predict the ratio $\frac{N_e}{N_\mu}$ as a function of shower depth. (ii) $X_{Max}$-tail: for fixed energy of the primary particle,  the probability of having a shower maximum deeper and deeper in the atmosphere decreases exponentially and therefore fitting the distribution of the position of shower maximum as a function of the depth allows to reconstruct $x_1$, Figure~\ref{fig:cos12} (right pane). To  select a sample of primary particles rich in protons the  deeper tail of the distribution is used in the fit, as heavier primary particles interact earlier on. The AUGER collaboration ~\cite{Collaboration:2012wt}, using this second method, has recently published a new result for the proton-air inelastic cross section at 57 TeV: $ \sigma^{57 TeV}_{Inel} (p-air) = 505\; \pm \; 22\; (stat) ^{+ 28} _{-36}\; (syst) \; mb.$ The range of values used in the fit   is chosen so that the  remaining helium and heavy nuclei contribute less than the statistical uncertainty. The first step of the measurement is the evaluation of $\Lambda_\eta$ which, using MC models, is linked to the value of $\sigma_{Inel}$. The determination of  $\Lambda_\eta$ has a systematic uncertainties of $\pm 15$ mb, while the remaining part of the systematic uncertainty is due to the process of extracting  $\sigma_{Inel}$. From the proton-air cross section, the values of the proton-proton inelastic and total cross-sections can be obtained using the Glauber model (for an introduction see ~\cite{Shukla:2001mb}) that basically describes the nucleus-nucleus interaction in terms of elementary nucleon-nucleon interaction. The Glauber model takes into account various nuclear and QCD effects such as nuclear geometry, opacity of nucleons, multiple interactions, diffraction, saturation, Fermi motion, the total, inelastic and elastic cross sections at lower energies and the value of the proton-air cross section at 57 TeV to provide an estimate of proton-proton total and inelastic cross sections. The AUGER collaboration finds that in the Glauber framework the inelastic cross-section is less dependent on model assumptions than the total cross-section. The result for the inelastic  and total proton-proton cross-sections are $\sigma^{57 TeV}_{Inel} (pp) = 92\; \pm \; 7\; (stat) \pm 9 \; (syst) \pm 7$ (Glaber)  mb  and $\sigma^{57 TeV}_{Tot} (pp) = 133\; \pm \; 13\; (stat) \pm 17 \; (syst) \pm 16$ (Glaber) mb. 
\vspace{-0.5cm}
\section{Collider experiments}
Collider experiments such as ALICE, ATLAS and CMS are able to directly measure the fraction of $\sigma_{Inel}(pp)$ composed by those events that leave {\it enough} energy in the  detector, typically in a rapidity interval $\eta \le |5|$, where the definition of {\it enough} is experiment and technique dependent. The results published so far rely on two different methods: (i) Forward energy (ii) Pile-up counting.
\vspace{-0.2cm}
\subsection{Pile-up method to determine $\sigma_{Inel}$}
\begin{wrapfigure}{!}{0.55\textwidth}
  \begin{center}
    \includegraphics[width=0.55\textwidth]{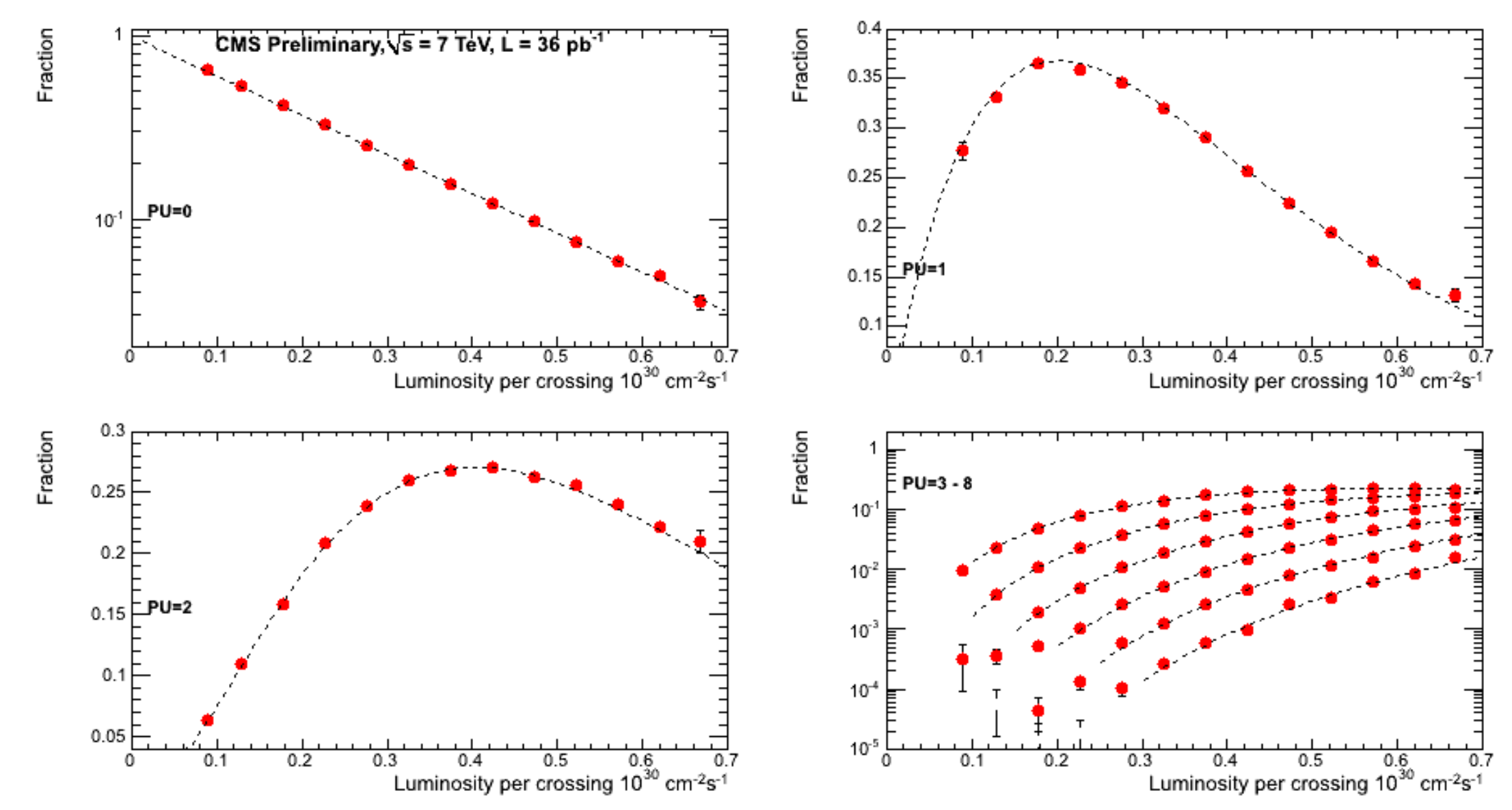}
  \end{center}
\caption{Fraction of $pp$ events with n pileup vertices, for n = 0 to 8, containing more than one track as a function of instantaneous bunch-crossing luminosity. The dashed lines are the fits using Equation~\ref{eq:pois}. }
\label{fig:pois}
\end{wrapfigure}
This method, used for the first time by the CMS collaboration~\cite{Chatrchyan:2012nj}, assumes that the number (n) of inelastic $pp$ interactions in a given bunch crossing follows the Poisson probability distribution:
\begin{equation}
\label{eq:pois}
P(n,\lambda) = \frac{\lambda^m e^{-\lambda}}{n!},
\end{equation}
\noindent where $\lambda$ is calculated from the product of the instantaneous luminosity for a bunch crossing and the total inelastic $pp$ cross section: $\lambda = L * \sigma_{Inel}$. The probability of having n inelastic $pp$ interactions, each producing a vertex with $>$1, $>$2, or $>$3 charged particles with $p_\perp > $ 200 MeV/c within $|\eta|$ = 2.4, for $n$ between 0 and 8, is measured at different luminosities to evaluate $\sigma_{Inel}$ from a fit of Equation~\ref{eq:pois} to the data. Figure~\ref{fig:pois} displays the data points as a function of the instantaneous luminosity, for events with n = 0 to 8 pileup vertices. For each n, the values of the Poisson distribution given	by Equation~\ref{eq:pois} are fitted as a function of $\lambda = L * \sigma_{Inel}$ to the data, providing an estimates of $\sigma_{Inel} (pp)$.
\vspace{-0.2cm}
\subsection{Forward energy method to determine $\sigma_{Inel}$}
The basic idea of this method is to count the number of events that, in a given interval of  integrated luminosity, deposit  at least a minimum amount of energy  in either of the forward part of the detector; the number of events is then converted into a value of cross section by accounting for detector  and pile-up effects.  ATLAS and CMS require to measure at least $E_{min} = 5$ GeV in the rapidity interval $3 \le |\eta| \le 5$ which is equivalent to, according to MC studies,  a minimum hadronic mass  $M_X$ of  at least 16  GeV/c$^2$ ( $\xi = M_X^2/s > 5 * 10^{-6}$).
\vspace{-0.2cm}
\subsection{Results}
\begin{figure}[h]
\begin{center}
 \resizebox{12cm}{!}{\includegraphics{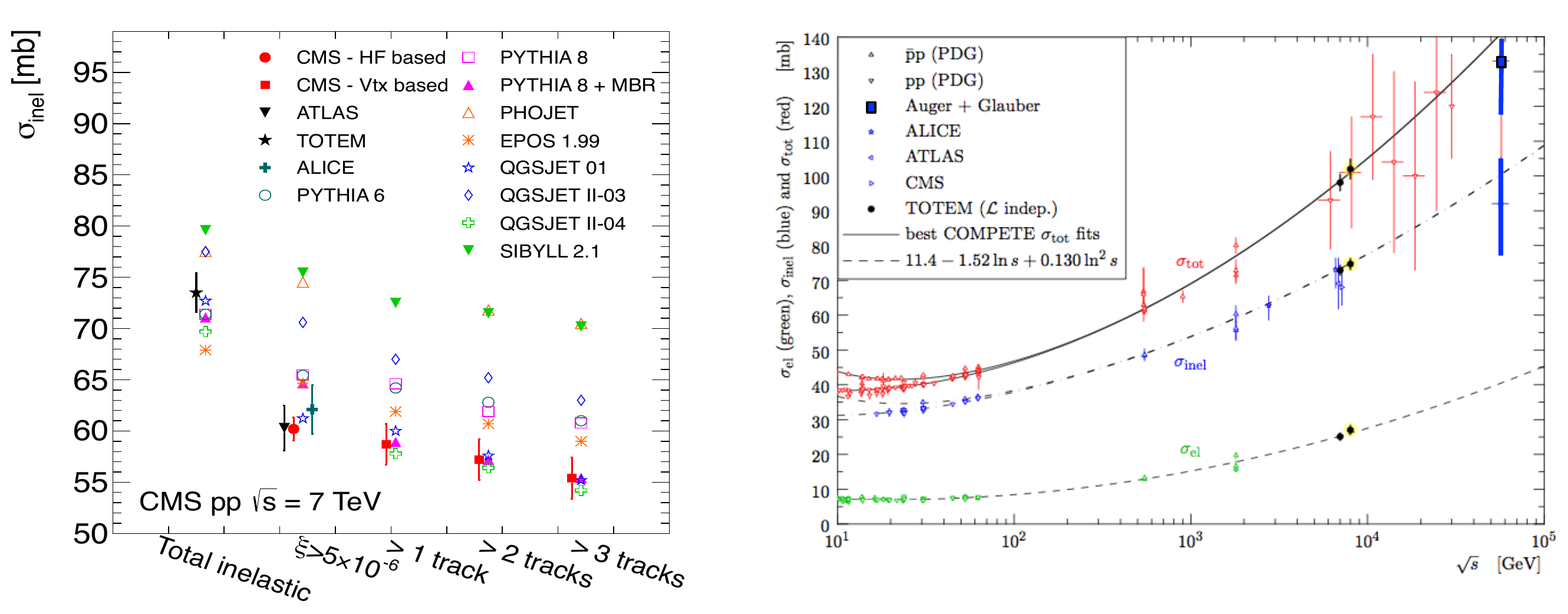}}
\end{center}
\caption{ Left pane: Compilation of the ALICE, ATLAS and CMS measurements of the inelastic $pp$ cross section  compared to predictions from several Monte Carlo models for different selection criteria, as labelled below the abscissa axis. The MC predictions have an uncertainty of 1 mb [11]. Right pane: compilation of the values of $\sigma_{Tot}(pp), \; \sigma_{El}(pp)$ and $\sigma_{Inel}(pp)$  as a function of the center-of-mass energy [8].}
\label{fig:res}
\end{figure}

\noindent Table~\ref{tab:all} lists the results from the ALICE~\cite{Abelev:2012sea}  ATLAS~\cite{Aad:2011eu} and   CMS~\cite{Chatrchyan:2012nj}  Collaborations for several selection criteria. The ALICE and ATLAS collaborations have extrapolated the measured values to provide also an estimate of $\sigma_{Inel}^{Tot}(pp)$. Figure~\ref{fig:res} (left pane) shows a compilation of the ALICE, ATLAS and CMS results for different selection criteria.  The data points are compared to a large set of MC models predictions, used both in cosmic-rays physics and collider experiments. Although several Monte Carlo models such as EPOS, QGSJET 01, QGSJET II-4, PYTHIA 6, and PYTHIA 8 reproduce correctly the value of $\sigma_{Tot}(pp)$, only QGSJET 01 and QGSJET II-04, and PYTHIA 8-MBR (but less	so)	are	able	to	simultaneously	reproduce	the	less	inclusive measurements. This observation suggests that most of the Monte Carlo models overestimate the contribution from high-mass events to the total inelastic cross section, and underestimate the component at low mass.
\begin{table}[h]
  \begin{center}
  \begin{tabular}{c c  c c c c } \hline
Exp &  Measurement  &  Result &  Stat & Syst & Lum \\ \hline
ALICE & $\sigma_{Inel}^{(\xi > 5\times 10^{-6})}$ &    62.1 &   & $  ^{+1.0}_{-0.9} $ & $ \pm 2.2$  mb \\
ATLAS & $\sigma_{Inel}^{(\xi > 5\times 10^{-6})}$ &    60.3 &$ \pm 0.05 $ & $ \pm 0.5 $ & $ \pm 2.1$ mb \\
CMS & $\sigma_{Inel}^{(\xi > 5\times 10^{-6})}$ &    60.2 & $ \pm 0.2 $&$  \pm 1.1 $&$ \pm 2.4 $ mb \\ \hline
ALICE & $\sigma_{Inel}$ &    73.2 &   & $  ^{+2.0}_{-4.6} $ & $ \pm 2.6$  mb \\
ATLAS & $\sigma_{Inel}$ &    69.4 &  & $ \pm 6.9 $ & $ \pm 2.4$ mb \\  \hline
CMS &  $\sigma_{Inel}^{({>}1\mbox{ track})}$ &  58.7 & & $ \pm 2.0 $&$ \pm 2.4 $ mb  \\
CMS &  $\sigma_{Inel}^{({>}2\mbox{ tracks})}$ & 57.2 & & $  \pm 2.0 $&$\pm 2.4 $ mb   \\
CMS &  $\sigma_{Inel}^{({>}3\mbox{ tracks})}$ &  55.4 & & $ \pm 2.0 $&$ \pm 2.4 $mb   \\
\hline
\end{tabular}
\caption{$\sigma_{inel}(pp)$ values for $\xi >  5\times 10^{-6}$ and for interactions with $>$1, $>$2 and $>$3 charged particles in the final state, with their uncertainties from systematic sources of the method and from luminosity.}
\label{tab:all}
\end{center}
\end{table}
\vspace{-0.5cm}

\vspace{-0.5cm}
 Figure~\ref{fig:res} (right pane) shows  a compilation of the values of $\sigma_{Tot}(pp), \; \sigma_{El}(pp)$ and $\sigma_{Inel}(pp)$  as a function of the center-of-mass energy.  The plots include the results obtained by the TOTEM collaboration, together with the results from LHC and lower energy experiments and the best fit from the COMPETE collaboration based on a  $ln^2(s)$ behaviour of the cross section.


\vspace{-0.2cm}
\section{Conclusions}
\vspace{-0.2cm}
The concurrent efforts of several cosmic-rays and collider experiments have provided in the last couple of years a large quantity of measurements on the values of the total, elastic and inelastic cross sections as well as the values of cross sections for particular final state. The total value of the cross section is very well reproduced by the prediction of the COMPETE collaboration and favours a  $ln^2(s)$ dependence of $\sigma_{Tot}(pp)$. Common MC models used in collider experiments fail to concurrently reproduce the new measurements, pointing to an underestimation of the amount of low mass events.

\end{document}